\documentclass[letter]{aa}  

\usepackage{graphicx}
\usepackage{txfonts}
\begin{document} 

\title{Addition of the Local Volume sample of galaxies from the FAST HI survey}
    \titlerunning{Addition of the Local Volume sample of galaxies from the FAST HI survey}
    \author{Igor D. Karachentsev\inst{1}, 
Valentina E. Karachentseva\inst{2}, 
 Serafim S. Kaisin\inst{1},
Chuan-Peng Zhang\inst{3,4}}

    \authorrunning{I.\,Karachentsev et al.}

    \institute{Special Astrophysical Observatory of the Russian Academy of Sciences, N.Arkhyz, KChR, 369167, Russia, idkarach@gmail.com,\and
Main Astronomical Observatory, National Academy of Sciences of Ukraine, Kiev, 03143, Ukraine, 
valkarach@gmail.com,
\and National Astronomical Observatories, Chinese Academy of Sciences, Beijing 1000101, People's Republic of China, cpzhang@nao.cas.cn,
 \and Guizhou Radio Astronomical observatory, Guizhou University Guyang 550000, People's
Republic of China}
    
 \abstract {We report the discovery of 20 new dwarf galaxies in the Local Volume identified as optical counterparts to the Five-hundred-meter Aperture Spherical radio Telescope (FAST) All Sky HI Survey (FASHI) sources. The galaxies have a median stellar mass of $7.8\times 10^6~M_{\odot}$ and  a  median HI mass of $1.0\times 10^7~M_{\odot}$. Most of them are field galaxies, while three are probable members of the M\,101 and M\,106 groups. We also found seven FASHI radio sources to be probable dark HI clouds in nearby groups. Together with four other known HI clouds in the local groups, their mean-square radial velocity difference of 49~km~s$^{-1}$ with respect to the host galaxies yields an average total mass  of ($2.7\pm1.0)\times10^{11}~M_{\odot}$ for these groups on the projected scale of 90 kpc.}



   \maketitle
%

\section{Introduction}

A Local Volume  (LV) of the Universe with a radius of 11 Mpc around the Milky Way is the most optimal sample 
for testing the results of $N$-body simulations performed within the framework of the standard cosmological 
paradigm, $\Lambda$ cold dark matter. A unique feature of this sample is  that it contains many faint dwarf galaxies that are essentially not observable at far distances. With distance and 
radial velocity measurements, the LV dwarf galaxies can be used as `test particles' to trace a local 
field of peculiar velocities and to recover a distribution of dark matter in the LV. The 
radius of the LV sphere, 
11 Mpc, is used because high-precision distances to  galaxies can be measured by  the  \textit{Hubble} 
Space Telescope from the tip of the red giant  branch (TRGB) in one orbital turnover of the telescope. 

The first sample of nearby galaxies within 10~Mpc was compiled by \citet{Kra1979} 
and 
included 179. Systematic all-sky searches for nearby low surface brightness galaxies  on images of 
the Palomar Observatory Sky Survey \citep{Kar1998,Kara2000,Kar1999} 
and subsequent 
measurements of their radial velocities \citep{Huc2000,Huc2001,Huc2003} 
increased the LV population to  approximately 500 objects. Surveys of large sky regions in  optical bands -- from the Sloan Digital Sky Survey \citep[SDSS;][]{Aba2009} and 
the Dark Energy Spectroscopic Instrument (DESI) Legacy Imaging Surveys \citep{Dey2019}  -- 
and in radio bands -- from the HI Parkes All Sky Survey \citep[HIPASS;][]{Kor2004} and 
the Arecibo Legacy Fast ALFA (ALFALFA) survey \citep{Hay2011} -- 
approximately doubled the number of  known  galaxies in the LV. Another significant addition to the LV sample resulted from special deep searches for  low surface  brightness objects in the virial zones of nearby   groups \citep{Chi2009,Mul2019,Car2022}. 
Summary data of the LV galaxies are presented in the Updated Nearby 
Galaxy Catalog  \citep[UNGC;][]{Kar2013}, 
a regularly updated version of which is available online\footnote {http://www.sao.ru/lv/lvgdb}. To date, the LV sample comprises about 1400 objects  with expected distances 
$D < 11$~Mpc.   

We note that `blind' sky surveys that use the HI 21~cm line have contributed significantly
to the population of known late-type dwarf galaxies with star formation. However, the most extensive HI surveys (HIPASS and
ALFALFA) cover predominantly the southern sky with declination Dec $< +38^{\circ}$. For the northern 
polar cap region, only one blind survey, of the Canes Venatici constellation, has been performed with the Westerbork radio 
telescope \citep{Kov2009}. 
The new capabilities of the Five-hundred-meter Aperture Spherical radio Telescope \citep[FAST;][]{Jia2020} 
promise to significantly reduce the existing asymmetry of the sky HI view.

According to \citet{Zha2024}, 
the FAST All Sky HI Survey (FASHI) will survey half the total sky
 (22,000 square degrees)  in the declination interval between $-14^{\circ}$ and $+66^{\circ}$. With a median sensitivity of
0.76~mJy$\cdot$beam$^{-1}$, a velocity resolution of 6.4~km~s$^{-1}$, and a beam size of $FWHM = 2.9\arcmin $, the
FAST telescope is capable of detecting significantly more radio sources than the 300 m Arecibo Telescope. 
The first release of the FAST survey covers close to 7600 square degrees. A list of 41741 sources with redshifts $z < 0.09$ detected in this survey has recently been published by \citet{Zha2024}. 

\section{New Local Volume candidates }

\begin{figure*}
\centering
\includegraphics[width=16.4cm,clip]{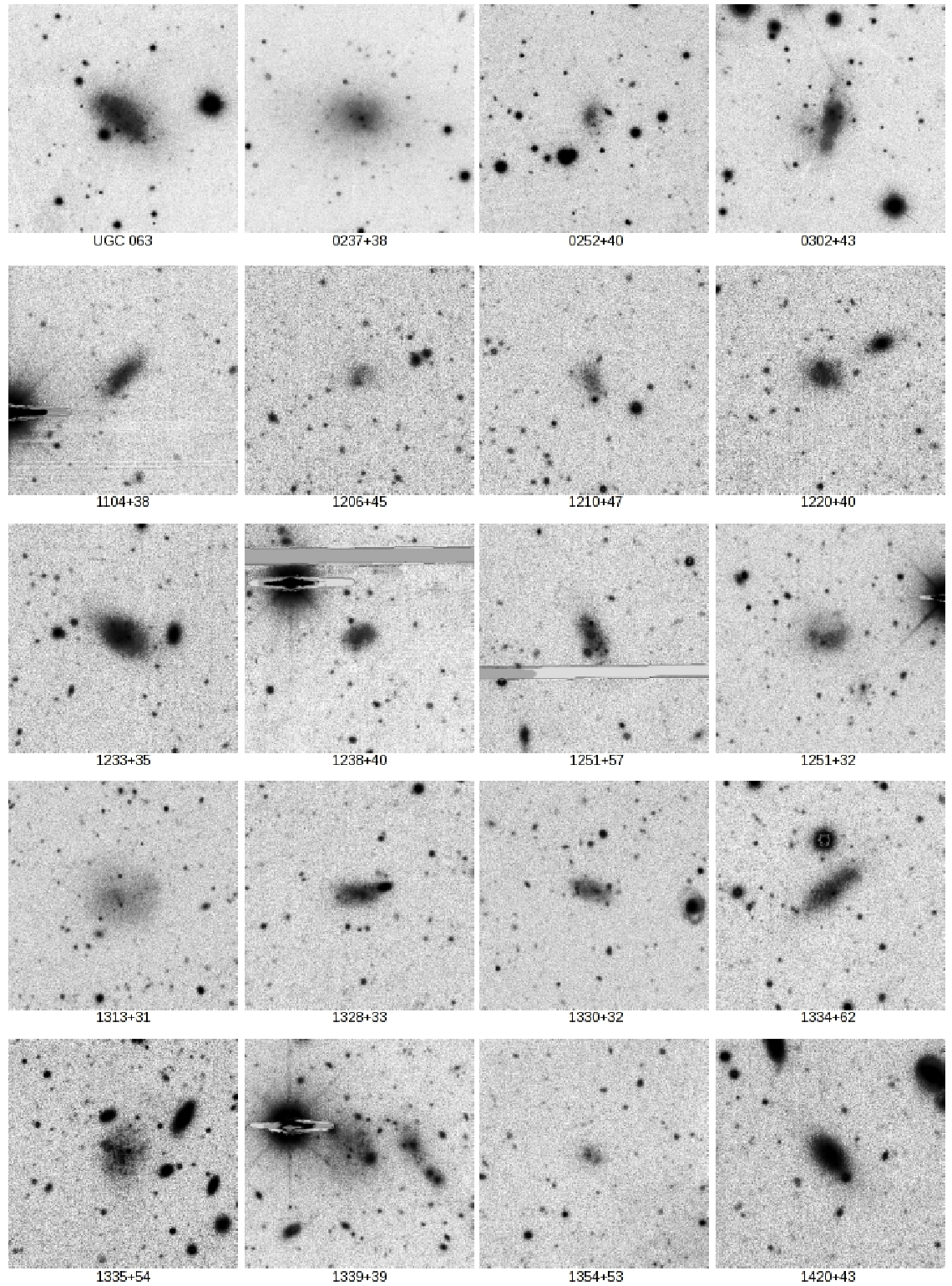}
\caption{Images of the 20 new dwarf galaxies identified with the FASHI radio sources. Most of the images are taken from the DESI Legacy Imaging Surveys. Each image side is $2\arcmin$. North is up, and east is to the left.}
\label{appfig1}
\end{figure*}

Among the radio sources from the \citet{Zha2024} catalogue, 
there are 580 objects with heliocentric radial velocities 
$V_{h}< 1000$~km~s$^{-1}$. Many of them are identified with well-known LV galaxies. Inspecting
this sample, we discovered 27 sources absent from the LV database. Seven of these radio sources are completely invisible in the DESI Legacy Imaging Surveys. They may be dark galaxies without any stellar counterpart or be equivalents to the high velocity clouds of the Milky Way.

\begin{table*}
\small
\caption{LV galaxy candidates from FASHI DR1.} \label{Table1}
\begin{tabular}{|lllcccrcccclll|}  \hline
  Name       &      RA  (2000.0)DEC &      $V_h$  &  $W_{50}$ &   $S_{HI}$  &   $D_{NAM}$ &  $m_{FUV}$ &   $g$     &   $r$   &  $B$ &$\log M_{\rm HI}$ &$\log M_*$     & $D_{TF}$   & Note\\\hline 
 \hline
    (1)      &       (2)            &  (3)   &    (4)   &     (5)   &    (6)      &      (7)    &   (8)      &   (9)      &  (10)    &   (11) &     (12)   &     (13)   &  (14)\\
\hline
UGC 63       &  001.962 +35.966        &    438   &     39    &    3.14     &     9.17    &    17.38       &    ---     &   ---  &   15.58  & 7.79 & 8.01 &     6.3   &  \\
FASHI0237+38 &  039.329 +38.930        &    420   &     34    &    0.76     &     9.76    &       ---      &      ---   &   ---  &   17.8   & 7.23 & 7.20 &    11.4  & \\
FASHI0252+40 &  043.042 +40.779        &    375   &     28    &    0.32     &     9.18    &       ---      &      ---   &   ---  &   19.0   & 6.80 & 6.70 &    13.8  & \\
FASHI0302+43 &  045.581 +43.663        &    289   &     21    &    0.43     &     8.15    &    19.31       &  16.97   &   16.53  &   17.33  & 6.83 & 7.38 &     5.8   & \\
FASHI1104+38 &  166.148 +38.737        &    610   &     24    &    0.40     &     9.64    &      ---       &  17.80   &   17.74  &   18.04  & 6.94 & 7.00 &     9.5   & \\
FASHI1206+45 &  181.688 +45.494        &    610   &     24    &    0.23     &     9.69    &    21.28       &  20.39   &   20.17  &   20.68  & 6.70 & 5.95 &       ---   &   (1)\\
FASHI1210+47 &  182.566 +47.574        &    408   &     23    &    0.44     &     7.06    &    20.91       &  18.89   &   18.68  &   19.18  & 6.71 & 6.30 &     8.2   &   (2)\\
FASHI1220+40 &  185.042 +40.878        &    520   &     21    &    0.21     &     7.26    &    21.02       &  18.50   &   18.09  &   18.85  & 6.41 & 6.44 &    10.9   & \\
FASHI1233+35 &  188.313 +35.734        &    817   &     35    &    0.92     &    10.32    &    17.98       &  16.76   &   16.53  &   17.06  & 7.36 & 7.44 &    10.0   & \\
FASHI1238+40 &  189.627 +40.863        &    644   &     22    &    0.24     &     9.21    &    18.90       &  17.93   &   17.75  &   18.21  & 6.68 & 6.89 &     9.0   &  (1)\\  
FASHI1251+57 &  192.767 +57.379        &    426   &     21    &    0.59     &     9.03    &       ---      &  17.76   &   17.59  &   18.04  & 7.05 & 6.95 &     7.6   &  \\    
FASHI1251+32 &  192.914 +32.181        &    849   &     19    &    0.43     &    10.15    &    19.77       &  18.60   &   18.49  &   18.86  & 7.02 & 6.72 &     8.0   &  \\    
FASHI1313+31 &  198.328 +31.415        &    801   &     20    &    0.44     &     9.69    &    21.09       &  18.03   &   17.70  &   18.36  & 6.99 & 6.88 &     8.3   &  \\    
FASHI1328+33 &  202.144 +33.147        &    754   &     33    &    1.31     &     9.60    &    18.39       &  17.53   &   17.38  &   17.80  &  7.45 & 7.08 &    9.1   & \\     
FASHI1330+32 &  202.538 +32.287        &    744   &     24    &    0.72     &     9.38    &    18.62       &  17.45   &   17.44  &   17.68  & 7.17 & 7.11 &     8.0   & \\
FASHI1334+62 &  203.578 +62.960        &    392   &     35    &    0.81     &     9.92    &    19.75       &  17.77   &   17.55  &   18.06  & 7.27 & 7.03 &    12.5   &  (1)\\
FASHI1335+54 &  203.982 +54.742        &    358   &     21    &    0.43     &     7.46    &      ---       &  18.85   &   18.74  &   19.11  & 6.75 & 6.35 &     8.1   &  (3) \\
FASHI1339+39 &  204.938 +39.135        &    682   &     25    &    0.97     &     9.96    &    18.65       &  17.45   &   17.51  &   17.68  & 7.35 & 7.16 &     8.3   &   \\
FASHI1354+53 &  208.530 +53.787        &    297   &     22    &    0.30     &     6.43    &       ---      &  19.73   &   19.35  &   20.07  & 6.46 & 5.84 &     ---     &  (3) \\
FASHI1420+43 &  215.051 +43.020        &    628   &     32    &    0.48     &    10.86    &    18.65       &  16.97   &   16.73  &   17.27  & 7.12 & 7.41 &    10.0   &    \\
\hline 
\multicolumn{14}{l}{Notes:  (1) an isolated dwarf; (2) a satellite of M\,106; (3)~a probable satellite of M\,101.}
\end{tabular}
\end{table*}

Images of 20  galaxies, identified with the FAST sources, are presented in Fig.~\ref{appfig1}. The basic parameters of these galaxies are 
listed in Table~\ref{Table1}: (1) the galaxy names, (2) their equatorial coordinates (in degrees; epoch J2000.0), (3) the heliocentric radial velocity (in km~s$^{-1}$), (4) the width of the HI line at half intensity 
from the maximum (km~s$^{-1}$), (5) the integrated HI line flux (in Jy$\cdot$km~s$^{-1}$), (6) the distance to the galaxy (in Mpc) determined by its radial velocity and taking the local velocity field calculated  
according to the numerical action method (NAM) into account  \citep[][]{Sha2017,Kou2020}, (7) 
the apparent UV  magnitude in the far-UV band from the Galaxy Evolution Explorer \citep[GALEX;][]{Mar2005,Gil2007}, (8-9) the apparent $g$ and $r$ magnitudes from the DESI Legacy Imaging Surveys \citep{Dey2019}, (10) the integral  apparent $B$ magnitude determined by the relation $B = g +0.313(g - r)+0.227$ as recommended by Lupton\footnote{https://www.sdss3.org/dr10/algorit hms/sdssUBVRITransform.php\\ \#Lupton2005}, or taken from NASA Extragalactic Database (NED)\footnote {http://www.ned.ipac.caltech.edu} when a galaxy is outside the DESI Legacy Imaging zone, (11) the galaxy hydrogen mass expressed as $\log(M_{\rm HI}/M_{\odot}) = 5.37 +  2\log(D_{\rm Mpc}) + \log(S_{\rm HI})$, (12) the galaxy stellar mass determined as
$ log(M_*/M_{\odot}) = 12.23 + 2 log(D_{Mpc}) - 0.4 B$
with the apparent B magnitude corrected for Galactic extinction (to estimate the galaxy stellar mass via its $V$-band luminosity, we used the relation
$M_*/M_{\odot} = 1.4 (L_V/L_{\odot}), $ as justified by \citealt{mcg2014}, 
and the average color $ (B - V)= +0.37 $  for
late-type dwarf galaxies; \citealt{mak1998}),  and (13) the galaxy distance determined by the Tully-Fisher (TF) relation between the HI line width
and the absolute $B$ magnitude,  $M_B = -19.99 - 7.27 (\log W_{50} - 2.5$) \citep{Tul2008}. For some of the gas-dominated galaxies, we used the magnitude $m_{21} = 17.4 - 2.5 \log (S_{\rm HI}$) instead of the $B$ magnitude. We also include notes about surrounding galaxies  (14).
As can be seen from Fig. 1, the optical diameters of the galaxies, being less than the beam size of $ FWHM=2.9\arcmin$, do not exceed $ 1\arcmin$.

We also include the galaxy UGC\,63 in Table 1. Its HI parameters were known previously, but it has not been considered before as a member of the LV.

 We note that, in general, the NAM distance estimations   agree well with  those determined using the TRGB distance estimate method. However, in specific sky regions close to the
radius of the zero velocity sphere of  the Virgo cluster, for example in the region of the Coma\,I group around NGC\,4278 \citep{Kar2011},
the difference between $D_{\rm NAM}$ and $D_{\rm TRGB}$ is large. For our LV member candidates, we therefore selected from the FASHI catalogue \citep{Zha2024} 
only galaxies for which the $D_{\rm NAM}$ estimates do not differ greatly from the $D_{\rm TF}$ estimates.

\section{HI clouds in the LV galaxy groups}

\begin{table*}
\small
\centering
  \caption{HI clouds in the LV groups.} \label{Table2}
\begin{tabular}{|lccrcrclcrr|}  \hline
  Name         &    RA(2000.0)DEC   & $V_h$ &$W_{50}$& $S_{HI}$  & $D_{\rm NAM}$ &  Host      &$D_{\rm host}$& $\Delta_V$& $R_p$& $M_{\rm HI}$\\ \hline
\hline
    (1)        &       (2)         &    (3)  &   (4) &   (5)  &   (6)    &  (7)       & (8)       & (9)      &   (10)     &  (11)\\        
\hline
FASHI1219+46a  &  184.822 +46.738   &  403  &    48  &    6.84  &      6.89 &   M 106 &   7.66   &    -44    &  76 & 7.88 \\
FASHI1219+46b  & 184.962 +46.503    & 391   &   35   &   3.56   &     6.57  &   M 106 &   7.66   &    -56    & 109 & 7.56 \\
FASHI1219+46c  &  184.969 +46.632  &  388  &    36  &    3.41  &      6.54 &   M 106 &   7.66   &    -59    &  92 &7.53  \\
FASHI1231+41   &  187.770 +41.141  &  629  &    36  &    6.16  &      9.10 &   NGC 4490 &   8.91   &     43    &  79 & 8.08\\
FASHI1243+32   & 190.846 +32.723   & 617   &   46   &   5.74   &     7.21  &   NGC 4631 &   7.35   &     34    &  41 & 7.84\\
FASHI1250+41   & 192.601 +41.520   & 328   &   31   &   0.32   &     4.39  &   M 94 &   4.41   &     20    &  32 & 6.16 \\
FASHI1251+41   & 192.814 +41.594   & 235   &   43   &   0.69   &     3.17  &   M 94 &   4.41   &    -73    &  37 & 6.21\\
\hline
HIJASS1021+68  & 155.251 +68.700   &  46   &   50   &   10.0   &     2.75  &   M 81     &   3.70   &     84    & 151 &7.25\\
CVnHI          & 185.181 +46.209   & 420   &   20   &   0.28   &     6.58  &   M 106 &   7.66   &    -27    & 152 & 6.45\\ 
M94HIn9        & 192.967 +40.291   & 298   &   20   &   0.23   &     3.80  &   M 94 &   4.41   &    -10    &  65& 5.89\\
GBT1355+54     & 208.711 +54.647   & 210   &   27   &   1.10   &     4.54  &   M 101    &   6.95   &    -30    & 151& 6.73\\
\hline
\end{tabular}
\end{table*}                                                                                                                      

Table~\ref{Table2} presents seven cases for which nearby HI sources from the FASHI catalogue are not identified 
with any optical counterpart in the $g$ and $r$ bands of the Legacy Imaging Surveys, nor in the UV (GALEX) or infrared (Wide-field Infrared Survey Explorer) wavebands. These sources are all located in well-known nearby galaxy groups. The first six columns of Table 2 are similar to those of Table~\ref{Table1}. Columns (7) and (8) present the
name and distance (in Mpc) from the main galaxy in  these groups. Columns (9) and (10) give the difference between the radial velocities (in km~s$^{-1}$) of the HI cloud and the central galaxy, as well as  the projected separation between them (in kpc), assuming that the 
cloud is at the same distance as the host galaxy. Column  (11)
provides the hydrogen mass of the clouds calculated as described for Table 1.
The bottom part of Table~\ref{Table2} presents the relevant 
data for four other known HI clouds in the LV groups: HIJASS1021+68 \citep{Boy2001}, 
CVnHI \citep{Kov2009}, 
M\,94-HIn\,9 \citep{Zho2023}, 
and GBT 1355+54 \citep{Mih2012}. 

\begin{table*}
\centering
\small
\caption{New HI data for the known LV objects.} \label{Table3}
\begin{tabular}{|lccccrcc|}  
\hline
 Name          &    RA(2000.0)DEC    & $V_h$& $W_{50}$& $S_{HI}$& $D_{NAM}$& $\log M_{\rm HI}$ & $\log M_*$ \\
\hline
               &   $^{\rm hh\, mm\, ss}$ $^{\circ\circ}$ '' ""  &km s$^{-1}$  & km s$^{-1}$    &Jy$\cdot$km s$^{-1}$  &  Mpc     &  $M_{\odot}$ & $M_{\odot}$    \\
\hline
MCG+06-27-017  &    12 09 56.4+36 26 07  &  325 &  48     &1.43     &  4.26  & 6.78 & 7.56    \\
LVJ1218+4655   &    12 18 11.1+46 55 01  &  396 &  62     &3.11     &  6.83  & 7.53 & 7.10  \\
SBS1224+533    &    12 26 52.6+53 06 19  &  291 &  31     &1.46     &  6.36  &7.14 & 7.29     \\
PGC4074723     &    12 40 29.9+47 22 04  &  523 &  20     &0.46     &  8.53  & 6.89 & 6.73  \\
dw1303+42      &    13 03 14.0+42 22 17  &  449 &  33     &0.25     &  6.48  &6.39& 6.37  \\  
Dw1311+4051    &    13 11 41.3+40 51 47  &  604 &  21     &0.46     &  9.32  & 6.97 & 6.36  \\
CGCG 217-018   &    13 12 51.8+40 32 35  &  565 &  65     &4.88     &  9.01   &7.97 & 7.97 \\
dw1313+46      &    13 13 02.0+46 36 08  &  388 &  27     &0.75     &  6.22  & 6.83 & 6.50   \\
CGCG 189-050   &    13 17 04.9+37 57 08  &  334 &  27     &2.14     &  4.20  &6.95 & 7.06  \\
PGC2229803     &    13 27 53.1+43 48 55  &  434 &  30     &0.41     &  6.55  &6.62 & 7.20     \\
MCG+08-25-028  &    13 36 44.8+44 35 57  &  488 &  30     &1.03     &  7.87   & 7.17 & 7.50    \\
LVJ1342+4840   &     13 42 20.1+48 40 57 &   434&   22    & 0.54    &   7.59 & 6.86 & 7.29  \\ 
\hline
\end{tabular}
\end{table*}

We also identified in the FASHI catalogue 12 LV galaxies with known optical radial velocities for which  HI parameters 
were absent. Their HI parameters are listed in Table~\ref{Table3}: (1) a galaxy name, (2) equatorial 
coordinates, (3--5) HI galaxy parameters from FASHI \citep{Zha2024}, 
(6) the galaxy distance (in Mpc) calculated via   the NAM \citep{Sha2017,Kou2020}, and 
(7-8) estimates of hydrogen and stellar masses. 

Appendix A contains the names of the galaxies presented in Tables~\ref{Table1}--\ref{Table3} and  the corresponding ID numbers of the radio sources from the FASHI  catalogue \citep{Zha2024}.
Appendix B shows the profiles of all the new LV HI sources in the same sequence as given in Tables~\ref{Table1}--\ref{Table3}.

    \section{Discussion and conclusion}
As seen from  Table~\ref{Table1}, all nearby FASHI sources have radial velocities $V_h < 850$~km~s$^{-1}$. They are star-forming
dwarf galaxies of morphological types Irr, Im, or BCD with a median color index of $g - r =0.21$. The ones that lie 
within the GALEX viewing area have significant fluxes in the far-UV band. 
More than half of the HI sources with optical counterparts have a hydrogen mass that exceeds their stellar mass. The gas-rich dIrr galaxy FASHI1206+45 has a hydrogen-to-stellar mass ratio of 5.6.

\begin{table}
\centering
\small
\caption{Average parameters of the HI sources.} \label{Table4}
\begin{tabular}{|lccccc|} \hline
Category & 
N & 
$\log\frac{M_{\rm HI}}{M_{\odot}}$ & 
s.d. & 
$\log \frac{M_*}{M_{\odot}}$ & 
$\log \frac{M_{\rm HI}}{M_*}$ \\ 
\hline
New\,LV & 20 & 7.00$\pm$0.08 & 0.34 & 6.89 & 0.11 \\
HI-clouds & 11 & 7.05$\pm$0.24 & 0.75 & $<$5.15 & $>$1.90 \\
Known\,LV & 12 & 7.01$\pm$0.12 & 0.46 & 7.08 & $-0.07$ \\ \hline
\end{tabular}
\end{table}

Table~\ref{Table4}  presents the average hydrogen and stellar masses for three categories of the FASHI sources: new LV dwarfs with an optical counterpart, HI sources without an optical counterpart, and the known LV dwarfs with new HI redshifts. There is no significant difference in the mean HI mass between the three subsamples. However, the HI clouds have a noticeably greater standard deviation (s.d.) in HI masses. The characteristic hydrogen mass of the new LV objects is $1.0\times10^7~M\odot$ with a typical ratio $ M_{HI}/M_* = 1.3 $. Assuming that HI sources with no visible optical counterpart have $g$ > 21.5 mag, we estimate their average hydrogen-to-stellar mass ratio to be more than 80.

Most of these dwarf galaxies are located in the general field, though some of 
them are very isolated  objects. Three dIrrs turned out to be peripheral members of 
known groups around M\,101 and M\,106 (also known as NGC\,4258).

Estimates of distances to galaxies in Table~\ref{Table1}, made according to the classic TF relation \citep{Tul2008} 
 or based on the galaxy's baryonic analogue under the assumption that the shape of a dwarf is described by a rotating ellipsoid with an axis 
ratio 
$b/a\simeq0.6$ \citep{Kar2017}, 
are in satisfactory agreement with kinematic distance estimates (NAM), accounting for the local velocity field.  The mean-square difference of the distance estimates is 2.1~Mpc, or 22\% 
at the average distance of these objects, $D_{\rm NAM} = 9.5$~Mpc .

Relatively isolated dark HI clouds in the LV groups,  presented in Table~\ref{Table2}, have parameters $W_{50}$ and $S_{HI}$
close to the parameters of usual dIrr galaxies. Their mean-square velocity difference  relative to the  
host galaxies in the groups amounts to <$\Delta V^2>^{1/2} = 49$~km~s$^{-1}$, and their average projected separation
from the  main galaxies, <$R_p$> , is 90~kpc. These clouds  can be considered test particles that characterise 
the groups' kinematics. The average virial mass of the groups is 
$M_v/M_{\odot} = 1.18\times 10^6 <\Delta V^2\times R_p$>, where $\Delta V$ is expressed in km~s$^{-1}$ and $R_p$ 
in kpc \citep{Kar2021}, 
and equals ($2.7\pm1.0)10^{11}~M_{\odot}$. The median $K$-band 
luminosity of the host galaxies in the groups under consideration is $6.2\cdot10^{10} L_{\odot}$, which is close 
to the Milky Way luminosity. For such luminous galaxies, the virial radius of their dark halo amounts to $R_v\sim300$~kpc. Consequently, the 
subsystem of HI clouds is concentrated predominantly in the central part of the groups, with a characteristic 
radius   three times less than $R_v$. In this volume, the ratio  $M_v/ L_K = (4.4\pm1.7)M_{\odot}/L_{\odot}$, estimated via the 
kinematics of HI clouds, is in line with the average ratio of <$M_v/L_K>  = 17.4\pm2.8(M_{\odot}/L_{\odot}$) 
obtained for halos of major spiral galaxies in the LV on the scale of their virial radius as based on the kinematics of satellites \citep{Kar2021}. 

It should be noted that for FASHI, the lower radial velocity limit  is $V_h = 200$~km~s$^{-1}$. 
However, there are about known 200 galaxies in the LV with radial velocities $V_h < 200$~km~s$^{-1}$. 
Hence, some fraction of the LV population will remain  undetected by this survey. Obviously, the establishment in 
FASHI of a lower radial velocity limit is due to the difficulty in detecting the HI signal of a galaxy against the bright local HI background from the Milky Way.

Our identification of radio sources from the first FASHI release with optical objects led to the detection of 20 new 
dwarf galaxy candidates in the  LV, as well as 7 HI clouds in nearby groups without signs of a stellar population. The upcoming completion of FASHI promises to increase the sample of LV galaxies.

The new LV candidates have radial velocities measured with a typical error of about 1~km~s$^{-1}$. Using new objects   
to trace the local field of peculiar velocities, determined by local attractors, requires
high-precision TRGB distances that need to be obtained with space telescopes. 
 
\begin{acknowledgements} The authors thank the referee for constructive comments
that helped to improve the paper. This work has made use of the FAST all Sky HI survey, the DESI Legacy
Imaging surveys, the NASA/IPAC Extragalactic Database (NED), the Galaxy Evolution Explorer (GALEX), and the revised version of the Local Volume galaxy database. IDK and SSK are supported by the grant 07--15--2022--262 (13.MNPMU .21.0003) of the Ministry of Science and Higher Education of the Russian Federation.
\end{acknowledgements}



\begin{appendix}
    
\section{Cross-identification with the FASHI catalogue}
  
\begin{tabular}{lc} 
\hline
Galaxy name     &       FASHI ID   \\
\hline
UGC 063         &    20230057537 \\
FASHI0237+38    &    20230057540 \\
FASHI0252+40    &    20230021207 \\
FASHI0302+43    &    20230025382 \\
FASHI1104+38    &    20230017803\\
FASHI1206+45    &    20230027281  \\
FASHI1210+47    &    20230029574  \\
FASHI1220+40    &    20230021454  \\
FASHI1233+35    &    20230058904  \\
FASHI1238+40    &    20230021429 \\
FASHI1251+57    &    20230038542  \\
FASHI1251+32    &    20230058517  \\
FASHI1313+31    &    20230007323  \\
FASHI1328+33    &    20230061148  \\
FASHI1330+32    &    20230008631 \\
FASHI1334+62    &    20230064037  \\
FASHI1335+54    &    20230036458  \\
FASHI1339+39    &    20230018394  \\
FASHI1354+53    &    20230063286  \\
FASHI1420+43    &    20230024668 \\
\hline
FASHI1219+46a   &   20230028812  \\
FASHI1219+46b   &   20230028556  \\
FASHI1219+46c   &   20230028794  \\
FASHI1231+41    &    20230021838 \\
FASHI1243+32    &    20230058573\\ 
FASHI1250+41    &    20230022318 \\
FASHI1251+41    &    20230022402 \\
\hline                           
MCG+06-27-017   &   20230013995   \\
LVJ1218+4655    &    20230028964  \\
SBS1224+533     &    20230034659  \\
PGC4074723      &     20230029371 \\
dw1303+42       &     20230023590\\
Dw1311+4051     &   20230021424   \\
CGCG 217-018    &   20230020782   \\
dw1313+46       &     20230028693 \\
CGCG 189-050    &   20230061751   \\
PGC2229803      &    20230025576 \\
MCG+08-25-028   &20230026440     \\
LVJ1342+4840    &   20230062806  \\ 
\hline
\end{tabular}                   
  
\section{HI profiles of the new Local Volume objects}

\begin{figure*}
\centering
\includegraphics[scale=0.80]{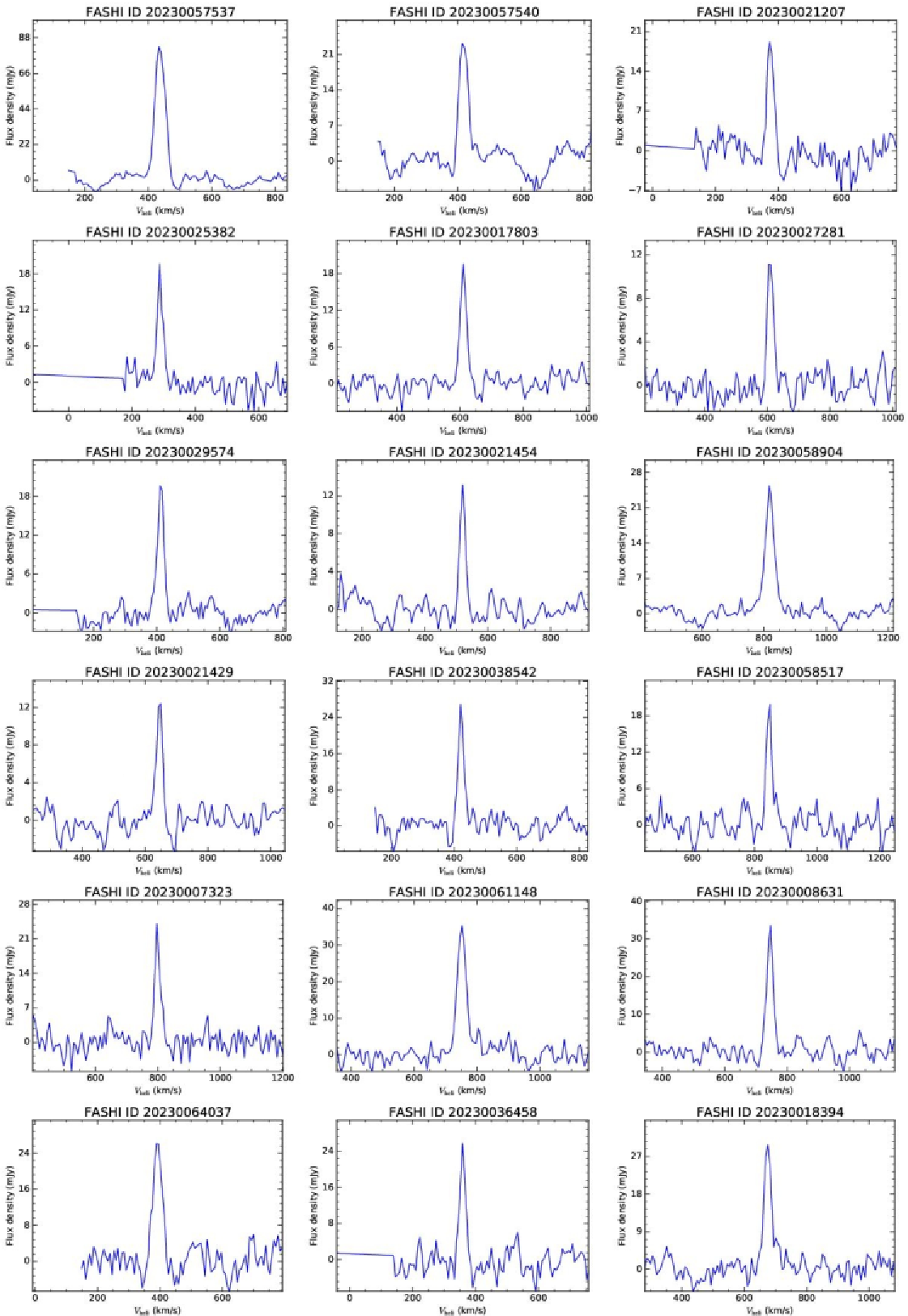}
\caption{Profiles of 39 new LV HI sources from FASHI given in the same sequence as in Tables~\ref{Table1}--\ref{Table3}.}
\end{figure*}


\begin{figure*}
\includegraphics[scale=0.8]{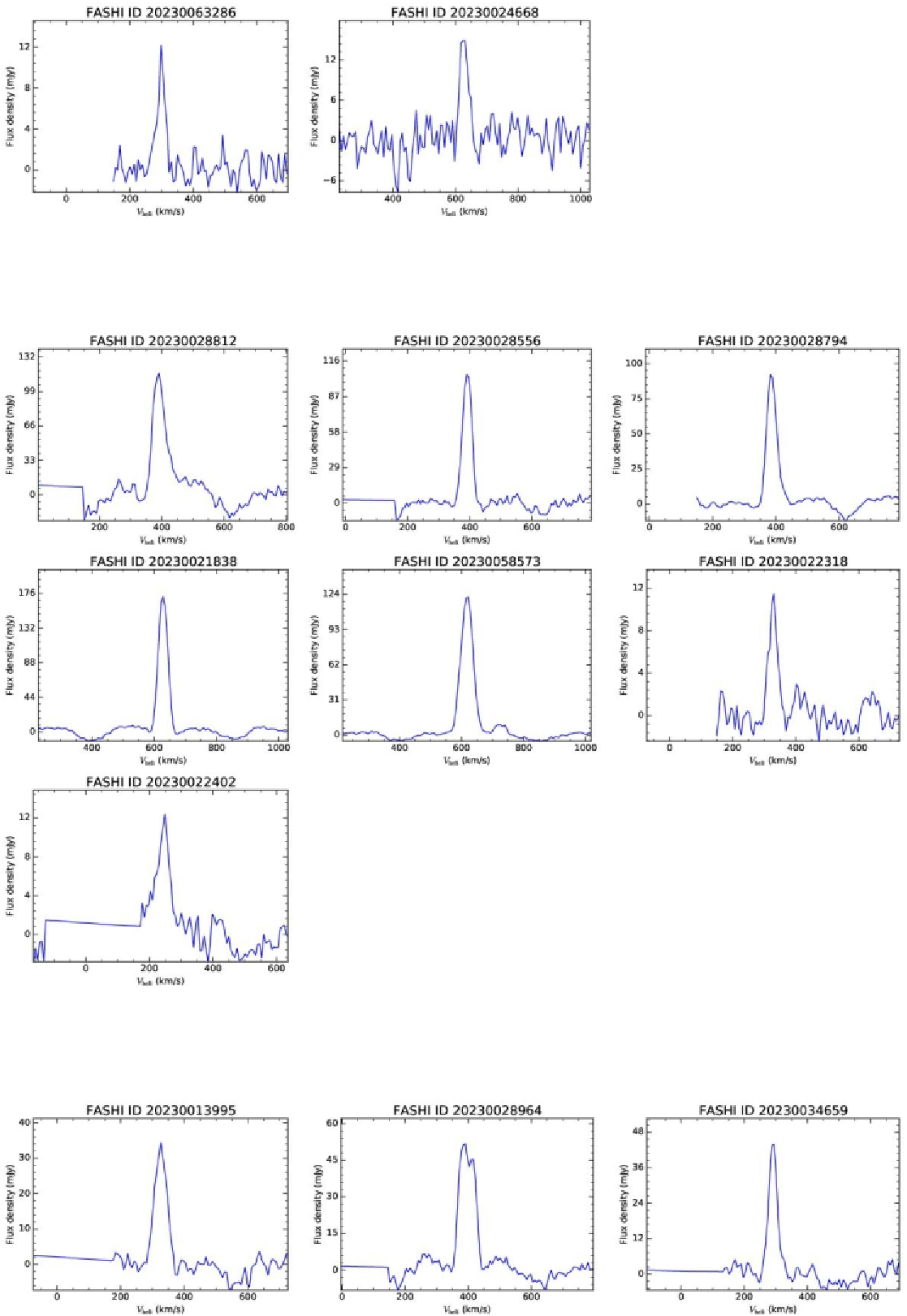}
\caption{Continued.}
\end{figure*} 
\begin{figure*}

\includegraphics[scale=0.8]{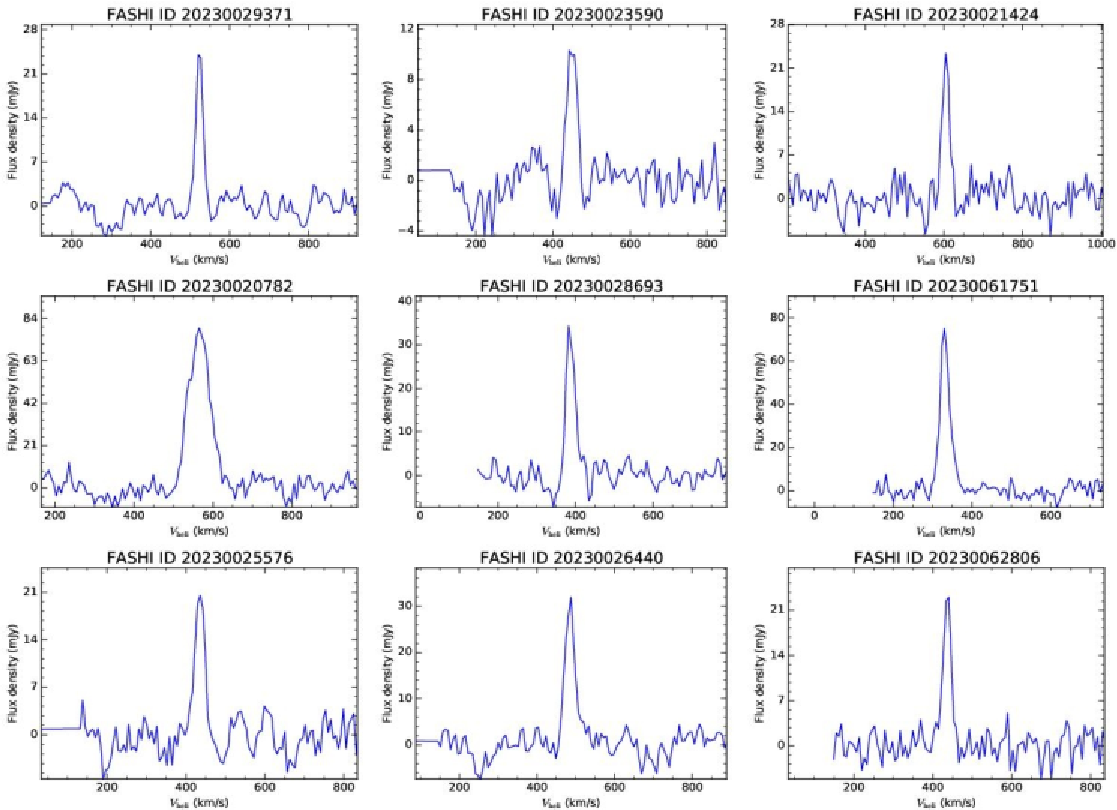}
\caption{Continued.}
\end{figure*}

\end{appendix}

\end{document}